\documentclass[12pt,a4paper]{article}
\pdfoutput=1

\usepackage{graphics}
\usepackage{epstopdf}
\usepackage{color}
\usepackage{ccaption}
\usepackage[labelfont={bf}]{caption}

\usepackage{amsfonts}
\usepackage{amssymb}
\usepackage{amsmath}
\usepackage{dsfont}
\usepackage{multirow}
\usepackage{latexsym}
\usepackage{verbatim}
\usepackage{mcite}
\usepackage{cite}
\usepackage{units}
\usepackage[all]{xy}
\usepackage[usenames,dvipsnames,table]{xcolor}
\usepackage{cancel}
\usepackage{slashed}
\usepackage{hyperref}
\usepackage{physics}
\usepackage{breqn}
\numberwithin{equation}{section}
\numberwithin{figure}{section}
\def\beq{\begin{equation}}
\def\eeq{\end{equation}}
\def\beqa{\begin{eqnarray}}
\def\eeqa{\end{eqnarray}}

\def\bfone{\relax{\rm 1\kern-.35em 1}}

\newcommand{\be}{\begin{equation}}
\newcommand{\ee}{\end{equation}}
\newcommand{\ben}{\begin{displaymath}}
\newcommand{\een}{\end{displaymath}}
\newcommand{\bea}{\begin{eqnarray}}
\newcommand{\eea}{\end{eqnarray}}

\newcommand{\bean}{\begin{eqnarray*}}
\newcommand{\eean}{\end{eqnarray*}}

\newcommand{\mc}[1]{\mathcal{#1}}

\newcommand{\parcial}[2][]{\frac{\partial #1}{\partial #2}}

\newcommand{\hs}[1]{\hspace{#1 cm}}
\newcommand{\la}{\langle}
\newcommand{\ra}{\rangle}
\newcommand{\riga}{\rightarrow}

\DeclareMathAlphabet{\mathpzc}{OT1}{pzc}{m}{it}

\begin{document}
\pagestyle{plain}

\begin{titlepage}
\begin{flushright}
%\small XXX
\end{flushright}

\bigskip

    \begin{center}
		{\Large \bf{Entanglement Renormalization of a $T\bar{T}$-deformed CFT
	}}
	
	\vskip 1.5cm

{\bf B.~Cardona$^\diamondsuit$\footnote{g.cardonarotger@um.es}\, and J. Molina-Vilaplana$^\clubsuit$\footnote{javi.molina@upct.es}}
	
\begin{center}
${}^\diamondsuit${\it Departamento de Electromagnetismo y Electr\'onica, Universidad de Murcia,}\\
{\it Campus de Espinardo, 30100 Murcia, Spain}

\vspace{0.5cm}

${}^\clubsuit${\it Departamento de Autom\'atica, Ingenier\'ia El\'ectrica y Tecnolog\'ia Electr\'onica, Universidad Polit\'ecnica de Cartagena,}\\
{\it Calle Dr. Fleming, S/N, 30202 Cartagena, Murcia}
\end{center}

\vspace{0.5cm}

	\today
	
\vspace{0.5cm}

    \end{center}

%%%%%%%%%%%%%%%%%%%%%%%%%%%%%%%
%%%%%%%%%%%%  ABSTRACT %%%%%%%%
%%%%%%%%%%%%%%%%%%%%%%%%%%%%%%%
\begin{abstract}
In this work we use cMERA, a continuous tensor network, to find a Gaussian approximation to the ground state of a $T\bar{T}$-deformed scalar CFT on the line, to first order in the deformation parameter. The result is used to find the correction to the correlators of scaling operators of the theory and to the entanglement entropy of a half-line. From the latter, we discuss the non-localities induced by the $T\bar{T}$ deformation at short length scales. We find that the kind of non-locality generated by those terms can be considered as a mild-one, in the sense that it does not violate the area law of entanglement. In the context of the conjectured connection between cMERA and holography, we find that at first insight a finite bulk radius can be defined in the putative geometric dual description of cMERA. However, the entropy analysis contradicts the proposal that no geometry can be ascribed to the region outside this radial cutoff.
\end{abstract} 

\end{titlepage}

%\pagenumbering{roman}
\tableofcontents
%\enlargethispage{1000pt}
%\pagebreak
%\pagenumbering{arabic}
%\setcounter{page}{1}

\section{Introduction}
\label{sec:intro}
Recently, there has been an increasing interest in studying irrelevant deformations of 2-dimensional conformal field theories (CFTs). In general, irrelevant operators lead to theories with non-renormalizability issues making them difficult to address. However, Refs.~\cite{Zamolodchikov:2004ce, Smirnov:2016lqw, Cavaglia:2016oda} identified a particular class of deformations built up from the bilinear product of the stress-energy tensor, and hence called $T\bar{T}$, that it turns out to be well-defined at arbitrarily large energy scales. As a result, in certain theories, the effect of the deformation on various physical observables can be computed non-perturbatively.

The $T\bar{T}$ deformation is defined by incrementally adding to the Lagrangian the $T\bar{T}$ operator, built up from the stress-energy tensor. More concretely, given a generic CFT in 2 dimensions with Lagrangian $\mathcal{L}_{0}$, the $T\bar{T} $-deformed theory is found by solving the flow equation \cite{Zamolodchikov:2004ce, Smirnov:2016lqw, Cavaglia:2016oda}:
\beq\label{eq:floweq}
\parcial[\mc{L}(\mu)]{\mu} = \mathcal{O}_{T\bar{T}}\,,
\eeq
subject to the boundary condition $\mathcal{L}(0) = \mathcal{L}_{0}$, where $\mu$ is the coupling or flow parameter, and the source is: \beq
\mathcal{O}_{T\bar{T}} = -\frac{1}{2}\left(T^{ab}T_{ab} - (T_a^a)^2\right).
\eeq
Notice that the stress-energy tensor entering in the r.h.s.~of the equation is that of the deformed theory, $T_{ab}(\mu)$, which relates to its trace via the so-called `trace relation'\,\footnote{This relation in fact, is true only inside an integral since the operator $\mathcal{O}_{T\bar{T}}$ is only defined up to derivatives. However, is correct at first order in $\mu$ since the operator product defining $\mathcal{O}_{T\bar{T}}$ is nonsingular when the stress tensor is that of a CFT.}\,: \beq\label{eq:tracerel}
T_a^a = 2\mu\, \mc{O}_{T\bar{T}}\,.
\eeq

This definition specifies the operator that is turned on at each point along the $\mu$-flow. In 2-dimensional field theory the deformation operator has dimension four, so the coupling has dimensions of $[\mu]=[{\rm length}]^2$. This defines a length scale, $\sqrt{\mu}$, at which the deformed CFT becomes non-local since the $T\bar{T}$ deformation is shown to generate an infinite number of higher-derivative interactions. However, the integrability property and/or UV completeness of $T\bar{T}$ is guaranteed. Beyond these very interesting properties, this kind of deformations have been seen to play an interesting role in the AdS/CFT correspondence \cite{McGough:2016lol}, which adds further value to their study. In particular, it was proposed that a $T\bar{T}$-deformed CFT is dual to 3-dimensional gravity in AdS with a radial cutoff, $r_c$. The holographic dictionary \cite{Maldacena:1997re, Gubser:1998bc, Witten:1998qj} provides the relation between the parameters in each side of the duality: $\mu = 16\pi G/r_c$ \cite{McGough:2016lol}. In this correspondence the holographic radial direction emerges from the $T\bar{T}$ flow parameter in the dual CFT. In typical states, the $T\bar{T}$-deformed observables coincide with those measured by bulk observers at the fixed radial position $r_c$ in AdS$_3$. 

In this work we study the entanglement renormalization  of a $T\bar{T}$-deformed CFT. That is to say, we analyze a $T\bar{T}$-deformed CFT in terms of a continuous MERA (a.k.a.~cMERA) tensor network \cite{Haegeman:2011uy}. cMERA consists of a variational real space renormalization group technique that starting from a simple and totally uncorrelated IR reference state, iteratively adds quantum correlations between small adjacent regions of space at different length scales. This flow eventually yields a UV approximation to the ground state of the theory under consideration. The original cMERA proposal provides exact results for free field theories \cite{Haegeman:2011uy, Nozaki:2012zj} and it is well understood.  Although new proposals extending cMERA for interacting field theories have appeared in the literature \cite{Cotler:2018ehb, Cotler:2018ufx, Fernandez-Melgarejo:2019sjo, Fernandez-Melgarejo:2020fzw, Fernandez-Melgarejo:2021mza}, our approach relies on the Gaussian cMERA setup.

Since the pioneering work of Swingle \cite{Swingle:2009bg}, it has been conjectured a connection between cMERA and the AdS/CFT. This is based on the entanglement structure of MERA-like tensor networks and that of the holographic states, as given by the Ryu-Takayanagi formula \cite{Ryu:2006bv}. In light of \cite{McGough:2016lol}, it is interesting to investigate how tensor networks conjectured to be dual to holographic geometries deal with the $T\bar{T}$ flow. Ref.~\cite{Kruthoff:2020hsi} explored in this direction and argued that a unitary representation of the $T\bar{T}$ flow in Eq.~\eqref{eq:floweq} can be interpreted as a superposition of continuous tensor network states. It was pointed out that a better understanding of this statement would lead to more explicit realizations of the AdS/tensor network correspondence. On the other hand, Ref.~\cite{Caputa:2020fbc} considered the $T\bar{T}$ flow deformation as ``folding in'' the Euclidean path-integral representation of the boundary CFT state towards the time-reflection symmetric slice in a dual bulk geometry. In that proposal, the different path integral representations of CFT states have a geometric dual given by a slice in the bulk, each one having the geometry of a hyperbolic space. Additionally, they interpret the states obtained from the Euclidean path integral of the $T\bar{T}$ flow deformed field theory as a family of continuous tensor network states. Despite no explicit constructions are given in the cited work, the interpretation of slices of a putative dual geometry as tensor network states was discussed since the inception of the holography/tensor network duality \cite{Swingle:2009bg}. Therefore, it is natural to think on explicit tensor network constructions based on cMERA, where each cMERA network would correspond to a concrete slice of the $T\bar{T}$ flow. In this regard, the authors of Ref.~\cite{Jafari:2019qns} used the path integral optimization approach \cite{Caputa:2017urj, Caputa:2017yrh} to find the time-slice of geometries dual to vacuum, primary, and thermal states in $T\bar{T}$-deformed 2-dimensional CFTs. 

In this paper we provide results on the UV structure of $T\bar{T}$ states and their holographic interpretation using the tensor network language. Our main goal is, within the reach of the approximations, to sharpen the connection between certain tensor network states and their geometric realization in the AdS/CFT correspondence. 

The rest of this paper is organized as follows. Section \ref{sec:ERTT} represents the core of this work. We first review the $T\bar{T}$-deformed CFT of a massless scalar and the (Gaussian) cMERA formalism in sections \ref{subsec:TTb} and \ref{subsec:gcMERA} respectively. The actual cMERA approximation to the ground state of the theory to first order in perturbation theory is found in section \ref{subsec:opt}\,. The rest of this section (section \ref{sec:anal}\,) is devoted to the analysis and discussion of our result. Those comprise \emph{i)} the analysis of the the IR limit of the cMERA approximation, including the scaling dimension of operators, \emph{ii)} the non-local features of the optimized cMERA state through the computation of the entanglement entropy (EE) and \emph{iii)} a discussion on the holographic interpretation of our results in light of the conjectured relation between cMERA and the AdS/CFT correspondence. We close in section \ref{sec:concl} summarizing and providing an outlook of our results. We include in appendix \ref{app:1} a detailed computation of the energy functional and its minimization procedure on which our work is based on.

\section{Entanglement Renormalization of a $T\bar{T}$-deformed CFT}\label{sec:ERTT}
\subsection{The $T\bar{T}$-deformed CFT}\label{subsec:TTb}
In this paper we discuss the simplest example of a $T\bar{T}$-deformed CFT: that one whose seed theory corresponds to a free boson. The (seed) CFT is then given by a free massless scalar field with Lagrangian: \beq
\mc{L}_{0} = \frac{1}{2}X\,.
\eeq
In $d = 2$ spacetime dimensions the invariant is $X \equiv (\partial\phi)^2 = (\dot{\phi})^2 - (\phi')^2$.\,\footnote{We use the notation $\dot{\phi}\equiv \partial_t\phi$ and $\phi'\equiv \partial_x\phi$.} For this theory, the $T\bar{T}$ flow equation (\ref{eq:floweq}) can be solved exactly. The Lagrangian is found to satisfy the equation $\partial\mc{L}/\partial\mu = -X^2\partial_X(\mc{L}^2/X)$, which can be recast into the form of Burger's equation \cite{Zamolodchikov:2004ce,Smirnov:2016lqw, Cavaglia:2016oda, Caselle:2013dra}. The solution is given by: \beq\label{eq:Lag}
\mc{L}(\mu) = \frac{1}{2\mu}\left(-1 + \sqrt{1+2\mu X}\right).
\eeq
The associated Hamiltonian at finite $\mu$ turns out to be: \beq\label{eq:Hmufinite}
H(\mu) = \frac{1}{2\mu}\int\dd x\left(1-\sqrt{1 - 4\mu\left(\mc{H}_0 - \mu P_0^2\right)}\right),
\eeq
where \beq
\mc{H}_0 = \frac{1}{2}\left(\pi^2+(\phi')^2\right)\,, \hs{0.75} P_0 = \pi \phi'\,,
\eeq
are the energy and momentum density of the seed CFT. The field $\pi$ is the canonical momentum conjugate to $\phi$. 

In this work we want to obtain a cMERA network representing the ground state of the theory with Hamiltonian density given by \eqref{eq:Hmufinite}. As we are considering an interacting theory, it is tempting to approach it in terms of  interacting cMERA circuits such as icMERA \cite{Fernandez-Melgarejo:2019sjo,Fernandez-Melgarejo:2020fzw}. Nevertheless, the icMERA circuit is devised to non-perturbatively address polynomial interactions such as the $\lambda\phi^4$-theory or the Gross-Neveau model. In this case, however, for finite values of $\mu$, the variational procedure of icMERA or even Gaussian cMERA seems rather involved due to the presence of a square root in the Hamiltonian. For this reason, here we proceed perturbatively. Indeed, it has been shown that a Gaussian ansatz given by a Gaussian cMERA state captures correctly the ground state at 1-loop in the $\lambda\phi^4$-theory \cite{Cotler:2016dha}. In the case of a $T\bar{T}$-deformed scalar field theory with non-zero mass \cite{Rosenhaus:2019utc}, the perturbative expansion of the Lagrangian to leading order in $\mu$ contains terms of the form $\mu (\partial \phi\, \bar{\partial}\phi)^{2}$ and $\mu m^{4} \phi^{4}$. Those are the kind of 1-loop terms that a Gaussian ansatz must grasp \cite{Barnes:1978cd, Stevenson:1985zy}. Bearing this in mind, we consider the $T\bar{T}$-deformed free boson \eqref{eq:Hmufinite} to leading order in $\mu$, yielding the perturbative Hamiltonian: \beq\label{eq:hamilt}
H_{T\bar{T}} = \int\dd x\left(\mc{H}_0+\mu\mc{O}^{(1)}_{T\bar{T}}(x)+ \mc{O}(\mu^2)\right)
\eeq
with \beq\label{eq:OTTmu}
\mc{O}^{(1)}_{T\bar{T}}(x) = \mc{H}_0^2 - P_0^2\,.
\eeq

\subsection{Gaussian cMERA}\label{subsec:gcMERA}
In this work we focus on cMERA descriptions of scale invariant scalar theories in $(1+1)$ dimensions. Here we briefly review the essentials of the cMERA formalism for these theories and refer the reader to Ref.~\cite{Haegeman:2011uy} and/or \cite{Nozaki:2012zj} for further details.

Given a Gaussian reference (IR) vacuum state, $|\Omega\ra$, defined via the annihilation condition \beq\label{eq:refvac}
\left(\sqrt{\frac{\Lambda}{2}}\phi(k) + \frac{i}{\sqrt{2\Lambda}}\pi(k)\right)|\Omega\ra = 0\,, \hs{0.75} \forall k\,,
\eeq
where $\Lambda$ is a UV cutoff, cMERA describes a continuous Hamiltonian entangling evolution in scale (labeled by $s$) of the quantum field degrees of freedom, from the IR scale to some UV length scale, \beq\label{eq:cMERAstate}
|\Psi(s)\ra = U(s_{\text{IR}},s)|\Omega\ra\,.
\eeq
The evolution is implemented through a continuous RG flow generated by the unitary: \beq
U(s_1,s_2) = \mathcal{P} \exp\left(-i\int_{s_1}^{s_2}\dd s\big(K(s)+L\big)\right).
\eeq
This unitary is a path-ordered exponential consisting of two (quasi-)local operators: an entangler $K(s)$ and a dilatation operator $L$; $s$ is the continuous RG scale coordinate, whose value increases when evolving from the reference state in the IR to the cMERA state in the UV. Typically one chooses $s_{\text{UV}} = 0$ and $s_{\text{IR}} = -\infty$. The dilatation operator $L$ is fixed by the matter content of the given QFT, whereas $K(s)$ contains all the variational freedom that it is actually optimized to determine the approximated ground state of the given theory.

For instance, for free scalar theories in $(1+1)$ dimensions, as the one we address here, the non-relativistic dilatation operator $L$ does not depend on the scale $s$ but it does on the scaling dimensions of the fields and it is given by
\beq
L=-\frac{1}{2}\int\, \dd x \left(\pi\left(x\phi'\right) + \left(x \phi'\right)\pi + \frac{1}{2}\left(\phi\,\pi + \pi\,\phi\right)\right).
\eeq

On the other hand, the entangler $K(s)$, in Gaussian theory, is given by the quadratic operator \beq\label{eq:entangler}	
K(s) = \frac{1}{2}\int\, \dd k\, g(k;s) \Big(\phi(k)\pi(-k)+ \pi(k)\phi(-k)\Big)\,,
\eeq
where $[\phi(k), \pi(k')] = 2\pi i\delta(k+k')\,$. The function $g(k;s)$, known as the density of disentanglers, is the only variational parameter. In fact, when evaluating the expectation value of the Hamiltonian of the theory on the cMERA state $|\Psi_{\Lambda}\ra \equiv |\Psi(s_{\text{UV}})\ra$, yields an energy functional in terms of the variational function\,\footnote{Note that for scale invariant field theories the density of disentanglers does not explicitly depend on the scale and thus $g(k;s) \equiv g(k)$.} \beq\label{eq:ftog}
 f(k,s_{\text{IR}})=\int^{s_{\text{IR}}}_0\dd s\, g(k e^{-s};s)\,,
\eeq
sometimes called the cMERA Bogoliubov angle (the name will come obvious shortly). Generically, in Gaussian cMERA one can express this latter function, once properly minimized, as \beq\label{eq:cMERAbangle}
f(k,s_{\text{IR}}) = \frac{1}{2}\log \frac{\epsilon(k)}{\Lambda}\,,
\eeq
with $\epsilon(k)$ the characteristic dispersion relation of the theory.

In terms of operators, the Gaussian cMERA flow implements an entangling evolution in scale or, equivalently, a continuous set of scale-dependent linear canonical transformations (a.k.a.~scale-dependent Bogoliubov transformations) on the fields\,\footnote{The interacting version of cMERA \cite {Fernandez-Melgarejo:2020fzw} implements instead a set of scale-dependent, \emph{non}-linear canonical transformations on the fields of the theory.} given by (in $d = 2$): \beq\begin{split}\label{eq:canotransf}
\phi(k) \hs{0.2} \riga \hs{0.2} \phi(k,s) &= e^{-f(k,s)}e^{-s/2}\phi(e^{-s}k)\,, \\
\pi(k) \hs{0.2} \riga \hs{0.2} \pi(k,s) &= e^{+f(k,s)}e^{-s/2}\pi(e^{-s}k)\,.
\end{split}\eeq
Notice that the transformations (\ref{eq:canotransf}) are generated through an entangler $K(s)$ that is at most quadratic in creation/annihilation operators. This creates optimized scale-dependent particle pair correlations, which is enough to account for the exact ground state of free field theories (since they are fully characterized by its 2-point correlators).

\subsubsection*{Magic cMERA}
We are interested in using a concrete realization of the Gaussian cMERA entangler known as \textit{magic} cMERA \cite{Zou:2019xbi}. For a free scalar field theory this is defined through a density of disentanglers profile given by:
\beq\label{eq:magic_mera}
g(k) = \frac{\Lambda^2}{2(k^2 + \Lambda^2)}\,.
\eeq
It has been shown that an optimized magic cMERA state $|\Psi_{\Lambda}\ra$ is the exact ground state of a \textit{strictly local} Hamiltonian (other known choices yield ground states of Hamiltonians that are quasi-local at best). That is, the state $|\Psi_{\Lambda}\ra $ is the exact ground state of the local Hamiltonian $H^{0}_{\Lambda} = H_0 + \Gamma^{\Lambda}_{\text{UV}}$, where $H_0 = \int\dd x\, \mc{H}_0$ is the free boson Hamiltonian and $\Gamma^{\Lambda}_{\text{UV}}$ is a non-relativistic UV-regulator that modifies it at small distances. The precise expression for this piece is: \beq\label{eq:UVreg}
\Gamma^{\Lambda}_{\text{UV}} = \frac{1}{2\Lambda^2}\, \int \dd x\, (\pi')^2\,.
\eeq

The optimization of the Gaussian cMERA state for the scalar field theory given by the Hamiltonian $H_\Lambda^0$ yields, following \cite{Haegeman:2011uy, Nozaki:2012zj}, a cMERA Bogoliubov angle of the form (\ref{eq:cMERAbangle}) together with the dispersion relation \beq\label{eq:cmera_bogo_dr}
\epsilon(k) = \frac{|k|}{ \sqrt{1+\left(k/\Lambda\right)^2}}\,.
\eeq
Therefore the UV-regulator $\Gamma^{\Lambda}_{\text{UV}}$ introduces $\mc{O}\left(|k|^3/\Lambda^2\right)$ corrections to the CFT dispersion relation $\epsilon_{\text{CFT}}(k) = |k|$, which are negligible at low energies.

\subsection{Optimization}\label{subsec:opt}
In this section we obtain the variational parameters of a Gaussian cMERA ansatz $|\Psi_\Lambda\rangle$ for the ground state of the Hamiltonian:
\beq\label{eq:h0hTT}
H_{\Lambda} = H_{T\bar{T}}+\Gamma^{\Lambda}_{\text{UV}}\,,
\eeq
where $H_{T\bar{T}}$ and $\Gamma^{\Lambda}_{\text{UV}}$ have been defined in Eqs.~(\ref{eq:hamilt}) and (\ref{eq:UVreg}) respectively. As noted above, the magic cMERA local Hamiltonian represents a scalar CFT modified at small distances by a non-relativistic UV-regulator. This has the effect of making the cMERA state $|\Psi_{\Lambda}\ra $ local at the range of distances at which the $T\bar{T}$ interaction precisely kicks in. This fact will be crucial for our results and it will be properly discussed later on.

In order to obtain the variational Bogoliubov angle $f(k,s)$ and the density of disentanglers $g(k,s)$, we compute the expectation value of the Hamiltonian \eqref{eq:h0hTT} w.r.t.~a generic Gaussian cMERA state. The result is given by: \beq\begin{split}
\frac{2}{\pi}\frac{\mc{E}[f]}{\text{vol}} &= \int\frac{\dd p}{2\pi}\Big(\Lambda e^{2f(p,s_{\text{IR}})}\left(1+\frac{p^2}{\Lambda^2}\right) + \frac{p^2}{\Lambda}e^{-2f(p,s_{\text{IR}})}\Big) \\
&\hs{0.45} +\frac{\mu}{4}\int\frac{\dd p}{2\pi}\frac{\dd q}{2\pi}\Big(3\Lambda^2e^{2(f(p,s_{\text{IR}})+f(q,s_{\text{IR}}))}+\frac{3}{\Lambda^2}p^2q^2e^{-2(f(p,s_{\text{IR}})+f(q,s_{\text{IR}}))} \\
&\hs{3}  -2q^2e^{2(f(p,s_{\text{IR}})-f(q,s_{\text{IR}}))} + 4pq\Big) + \mc{O}(\mu^2)\,,
\end{split}\eeq
where $\text{vol} \equiv \delta(0)$. Imposing this functional to be on a minima and solving the resulting equation for the variational function $f(k,s_{\text{IR}})$, one eventually finds: \beq \label{eq:f_bogoliubov}
f(k,s_{\text{IR}}) = \frac{1}{2}\log \frac{|k|}{\Lambda\sqrt{1+\left(k/\Lambda\right)^{2}}}\, \beta(k;\mu)\,,
\eeq
where \beq
\beta(k;\mu) = 1+\frac{a}{3 \pi}\left(\frac{1 + b \left(k/\Lambda\right)^{2}}{1+\left(k/\Lambda\right)^{2}}\right)\Lambda^2\mu\,,
\eeq
and $a= 2-\sqrt{2}$, $b=3\sqrt{2}/(4a)$. More details about the derivation of the energy functional and the Bogoliubov angle (\ref{eq:f_bogoliubov}) can be found in Appendix \ref{app:1}\,. It is worth stressing here that the result above is derived assuming $\Lambda^2\mu \ll 1$, which implies that the effect of the deformation is small even for momenta close to the cMERA cutoff.

\begin{figure}[t!] 
\begin{center}
\begin{minipage}{\textwidth}
\begin{center}
\includegraphics{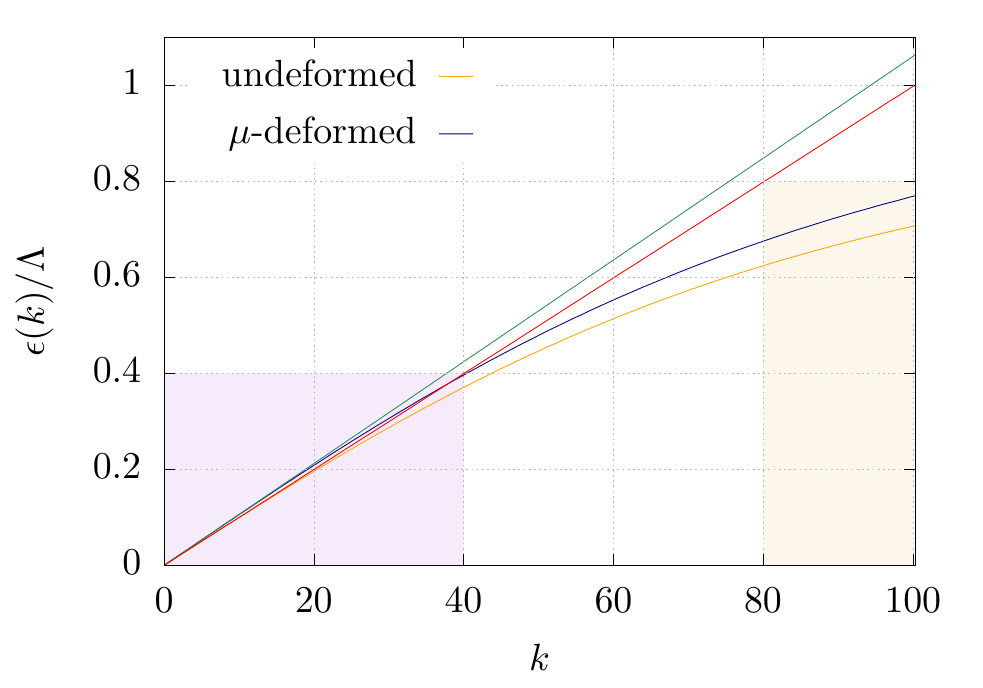}
\end{center}
\end{minipage}
\captionsetup{width=0.9\textwidth}
\captionof{figure}{\textsl{Dispersion relation $\epsilon(k)$ normalized with respect to the cutoff $\Lambda$ in Eq.~\eqref{eq:dispersion_relation} for the $T\bar{T}$-deformed theory vs the undeformed one. The straight lines represent the dispersion relation in Eq.~\eqref{eq:dispersion_relation_cl} for the deformed (green) vs the undeformed (red) theory.  The limit $k \ll \Lambda$ is represented as a purple shadowed region while the UV limit where the deformation acts is marked by the yellow-shadowed region. To perform this plot we took $\mu = 10^{-4}$ and $\Lambda = 1/\sqrt{\mu}$.}}
\label{fig:dispersion_rel}
\end{center}
\end{figure}

In order to obtain the density of disentanglers of the cMERA we just recall \eqref{eq:cMERAbangle}. According with \eqref{eq:f_bogoliubov}, the dispersion relation $\epsilon(k)$ of the $T\bar{T}$-deformed theory reads: \beq\label{eq:dispersion_relation}
\epsilon(k)=\beta(k;\mu)\, \epsilon_0(k)\,,
\eeq
where $\epsilon_0(k)$ is the dispersion relation of the undeformed theory, i.e.~that obtained for $\mu=0$ which has been defined in Eq.~\ref{eq:cmera_bogo_dr}\,. The comparison between those quantities is displayed in Fig.~\ref{fig:dispersion_rel}\,. We are now in a position for obtaining the density of disentanglers $g(k)$, which follows from the definition \cite{Nozaki:2012zj}: $g(k)=\frac{1}{2}k\, \partial_k\, \log\epsilon(k)$. In our case we obtain: \beq\label{eq:density_disentanglers}
g(k) = g_0(k) \left(1+  \frac{\alpha}{3 \pi}\, g_0(k)\, \mu\, k^2\right),
\eeq
with $g_0(k)$ being the density of disentanglers of the undeformed theory given in Eq.~\eqref{eq:magic_mera} and $\alpha = 7 \sqrt{2}-8$. The resulting density of disentanglers clearly shows that the deformation effect of the interacting $T\bar{T}$ term occurs for momentum modes such that $k\gtrsim 1/\sqrt{\mu}$, i.e.~at the UV length scale fixed by the deformation parameter $\mu$. This implies that our solution should recover a sensible undeformed IR limit for $k\ll 1/\sqrt{\mu}$. This is discussed in \S\ref{sec:anal}.

\subsubsection{Sensitivity to the cMERA regularization}
We would like to stress that using the magic cMERA regularization scheme is neither an arbitrary or capricious selection in this problem. As commented above, other realizations of the Gaussian cMERA entangler have been previously used \cite{Haegeman:2011uy,Nozaki:2012zj}. Optimization procedures with these realizations have shown to yield ground states that are the exact ground states of Hamiltonians that are quasi-local at best. The most common regularization scheme for the entangler is given by the factorized form: 
\beq
\label{eq:hard_cutoff}
g(k;s)=\chi(s)\Gamma(k/\Lambda)\,,
\eeq
where $\Gamma(x)\equiv \Theta(1-|x|)$ is the Heaviside step function. $\chi(s)$ is a real-valued function  and $\Gamma(k/\Lambda)$ implements a high-frequency sharp cutoff. Concretely, when optimizing the free massless scalar under this choice, one obtains $\chi(s) = 1/2$ and $f_0(k,s_{\text{IR}})=\frac{1}{2}\log\frac{|k|}{\Lambda}$. 

Despite this is a perfectly acceptable solution for the free scalar, the cMERA state obtained turns out to be the true ground state of a quasi-local Hamiltonian, especially at short distances. Conversely, the magic cMERA parent Hamiltonian is local at the range of distances at which the $T\bar{T}$ deformation is expected to act. In this sense, we carried out the optimization over $\mc{E}[f]$ in terms of the sharp cutoff cMERA realization obtaining: \beq\label{eq:eqEeq1}
f(k, s_{\text{IR}}) = \frac{1}{2}\log \frac{|k|}{\Lambda}\left(1+\Big(\frac{1}{\Lambda}I_{(2)}^- - \Lambda I_{(0)}^+\Big)\left(\mu + \mc{O}(\mu^2)\right)\right).
\eeq
Proceeding analogously as before, now the trivial dispersion relation at $\mu = 0$, $\epsilon_0(k) = |k|$, yields: $I_{(0)}^+ = I_{(2)}^-/\Lambda^{2} = \Lambda/2\pi$, resulting in the trivial solution $f(k, s_{\text{IR}}) = f_0(k, s_{\text{IR}})$.

In other words, the sharp cutoff realization of the cMERA disentangler cannot see the effect of the  $T\bar{T}$ interaction at first order in perturbation theory. We have checked that this is also the case at second order (see Appendix \ref{subsec:appA1}), and presumably at any order, in perturbation theory, making manifest the relevance of the cMERA regularization when dealing with irrelevant deformations that perturb the UV structure of the ground states.

\subsection{Entanglement Entropy and AdS/cMERA correspondence}\label{sec:anal}
This section is devoted to analyze and discuss several relevant aspects of the cMERA representation of the $T\bar{T}$-deformed theory just obtained. We start by taking its low-energy limit and then we move on to discuss what lessons we can learn from the point of view of entanglement entropy and holography.

\subsubsection{The IR limit}
The IR low-energy limit consists of taking $k\ll \Lambda$. In this limit the variational Bogoliubov angle \eqref{eq:f_bogoliubov} and the dispersion relation (\ref{eq:dispersion_relation}) take the following form: \beq\begin{split}\label{eq:dispersion_relation_cl}
f(k,s_{\text{IR}}) &= \frac{1}{2}\log\frac{|k|}{\Lambda}\beta(\mu)\,, \\
\epsilon(k)&= \beta(\mu) |k|\,,
\end{split}\eeq
 where $\beta(\mu)\equiv 1+(a/3 \pi)\Lambda^2\mu$. Taking into account that the cMERA IR dispersion relation for the undeformed theory is $\epsilon_0(k)=|k|$ (at low energies the UV-regulator does not contribute), we note that for $k\ll \Lambda$, the effect of the $T\bar{T}$ deformation simply amounts to  re-scale the undeformed dispersion relation. We give an interpretation of this re-scaling in section \ref{subsec:holo} in terms of the proposed connection between tensor networks and AdS/CFT.

On the other hand, the density of disentanglers in this limit reads: \beq\label{eq:density_disentanglers_cl}
g(k)=\frac{1}{2} \left(1+  \frac{\alpha}{6 \pi} \mu k^2\right),
\eeq
and therefore $g(k)\simeq 1/2$ only when $k\to 0$. This last result together with the equations of the cMERA renormalization flow for the fields (or 2-point functions) of the theory have an interesting consequence. The cMERA renormalization group flow equations for the operators of a scalar theory are given by \beq \label{eq:scaling_dimensions}\begin{split}
\partial_s \phi(k,s)&=-i\left[L + K(s),\phi(k,s) \right]=-\left(k\partial_k +\frac{1}{2}-g(k)\right)\phi(k,s)\\
&=-\left(k\partial_k +\Delta^{\phi}\right)\phi(k,s)\,, \\
\partial_s \pi(k,s)&=-i\left[L + K(s),\pi(k,s) \right]=-\left(k\partial_k +\frac{1}{2}+g(k)\right)\pi(k,s)\\
&=-\left(k\partial_k +\Delta^{\pi}\right)\pi(k,s)\,.
\end{split}
\eeq
Using \eqref{eq:density_disentanglers} we obtain: \beq\begin{split}
\Delta^{\phi,\pi} &= \frac{1}{2} \mp g(k) \\
&=\Delta_{0}^{\phi,\pi} \mp \frac{\alpha}{3 \pi} g^2_0(k) \mu k^2
\end{split}\eeq
with $\Delta_{0}^{\phi,\pi} = 1/2 \mp g_0(k)$. Taking the limit $k\ll \Lambda$ we get: \beq
\Delta^{\phi,\pi}= \Delta_{0}^{\phi,\pi} \mp \frac{\alpha}{12 \pi} \mu k^2
\eeq
with $\Delta_{0}^{\phi,\pi}=\lbrace{0,1\rbrace}$ respectively. This result implies that in the IR the 2-point correlators follow the power-law \beq
\mc{C}(k) \sim k^{2\Delta_{0} + 2\gamma \mu k^2}\,,
\eeq
 where $\gamma=\mp\alpha/12 \pi$ and $\Delta_{0}$ is the IR conformal dimension. This  indicates that the cMERA entanglement renormalization flow mixes a given operator $\mc{O}$ with $\nabla^{2}\mc{O}$ in a range of length scales running from those for which $k \geq 1/\sqrt{\mu}$ up to the UV scale $\Lambda$, and suggests the definition of a new kind of cMERA renormalized non-local operator $\mc{O}_R = \mc{O} + \gamma\mu \nabla^{2}\mc{O}$. Namely, we see how a quantity related to the scaling dimension of a local operator $\Delta_{\mc{O}}$ depends on a range of scales due to its $k^{2}$ dependence fixed by the coupling $\mu$. We would like to emphasize here the clear resemblance between our result and those obtained by Cardy \cite{Cardy:2019qao} for the Fourier-transformed renormalized 2-point functions obtained by solving the Callan-Symanzik equation. 

\subsubsection{Entanglement entropy and the non-locality of $T\bar{T}$}
One of the most known and discussed features of a $T\bar{T}$-deformed CFT is its non-locality for distances shorter than the scale fixed by the deformation parameter $\sqrt{\mu}$. In this section we discuss about the kind of non-locality that arises in our cMERA approximation. 

For a half-space bipartition of the total space, the entanglement entropy of a state approximated by a Gaussian cMERA is given by \cite{Fernandez-Melgarejo:2021ymz} \beq \begin{split}
S_A &= \frac{1}{6}\int \dd k\, \log \langle \Psi_\Lambda |\phi(k)\phi(-k)| \Psi_\Lambda \rangle \\
& = -\frac{1}{3} \int \dd k\, f(k,s_{IR}) + \frac{1}{6}\, \int \dd k\, \log \frac{\delta(0)}{2 \Lambda} \\
&= -\frac{1}{3}\int \dd k \int_0^{s_{\text{IR}}} \dd s\, g(ke^{-s}) + \widetilde{\kappa}_{\Lambda}\,,
\end{split}
\eeq
where $\kappa_\Lambda = \frac{1}{6}\, \int \dd k\, \log \frac{\delta(0)}{2 \Lambda}$. Given the structure of the density of disentanglers for the deformed theory given in Eq.~\eqref{eq:density_disentanglers}, it is straightforward to show that \beq\label{eq:ee_cmera_tt}
 S_A(\mu)  = \frac{\pi}{24}\Lambda\left(1 + \frac{\alpha}{12 \pi^2}\Lambda^2\mu\right) + \widetilde{\kappa}_{\Lambda}\,.
\eeq
Since $\Lambda^2\mu$ is dimensionless and, as stated above, the dispersion relation \eqref{eq:dispersion_relation} for the deformed theory is valid in the regime $\Lambda^2\mu \ll 1$, it is sensible to interpret that the non-locality of the deformed theory manifests itself merely as a correction to the area law in terms of the length scale $\sqrt{\mu}$. This interpretation is actually reinforced by the fact that a strong violation of the area law is associated to highly non-local dispersion relations (see below). In view of this, the non-locality of the deformed theory captured by our cMERA construction might be considered as a mild non-locality, insomuch as the resulting entanglement entropy does not violate the area law. Notice however that this is not trivial: since the cMERA Bogoliubov angle $f(k,s)$ \eqref{eq:f_bogoliubov} does include contributions from terms that break conformal symmetry, one would expect violations of the area law \cite{Shiba:2013jja}. 

As a fact, the well-known area law for the entanglement entropy comes as a direct consequence of the locality of the action of the theory. When non-local interactions are introduced in the action or Hamiltonian density, deviations from the area law are expected to appear. Nevertheless, as shown in \cite{Shiba:2013jja, Basa:2019ywr} the appearance of these putative violations is quite subtle. In Ref.~\cite{Basa:2019ywr} the authors considered two types of non-local actions denoted by \beq\begin{split}
I_B(\phi)&= \int\dd^{d+1} x \, \phi(x)(-\Delta)^w \phi(x) \, , \\
I_{C}(\phi)&= \int\dd^{d+1} x \, \phi(x)e^{ (-\Delta)^w}\phi(x)\, ,
 \end{split}
 \eeq
 with $w \in {\mathbb R}$, and $(-\Delta)^w$ is the fractional Laplacian. Remarkably, while both types of theories contain non-local operators, they show a completely different scaling behavior for the entanglement entropy. For a half-space, the relevant case in this discussion, one has:
\beq\label{eq:entropy_scalings}\begin{split}
S^{B}&\sim c_{d-1}\tilde{\Lambda}^{d-1}+\dots\,,\\
S^{C}&\sim c_{d-1}\tilde{\Lambda}^{d-1+2w}+\dots
\end{split}
\eeq
Here $\tilde{\Lambda}$ is a momentum cutoff that acts as the inverse of a short-distance cutoff and $c_{d-1}$ is a constant. It is clear that in $B$-type theories the entanglement entropy follows the prototypical area law of a local QFT. This is the kind of theory we have referred to before as having mild non-localities. On the other hand, $C$-type theories deviate strongly from the area law behavior and have a non-local dispersion relation \cite{Shiba:2013jja} \beq
 \epsilon_{C}(k)\propto \exp\left(k^{2 w}\right), 
 \eeq
 and, in the particular case of $w=1/2$, can yield a volume law \cite{Shiba:2013jja}. For the free scalar theory defined by $I_{C}(\phi)$, a cMERA Gaussian ansatz would yield a density of disentanglers profile of the form \beq g_C(k)= w k^{2 w}\, .
\eeq
We thus see that the density of disentanglers in Eq.\eqref{eq:density_disentanglers} is not of the C-type. Indeed, for $k \to \Lambda$, $g(k)\to g_0(\Lambda) \sim \mathcal{O}(1)$ as our cMERA construction is valid for $\Lambda^2\mu \ll 1$.
  
Summarizing, not all non-localities in the action might give rise to deviations from area law. In line with this discussion, it worths to mention that in Ref.~\cite{Basa:2019ywr} the authors showed that it is possible to recast all $B$-type theories via a Hilbert space transformation as purely local theories. They also concluded that $C$-type theories are just extreme cases for the non-applicability of this reduction, and conjectured that the existence of a local Hilbert space transformation indicates whether the action for a QFT is truly local. We leave for future research to explore whether there exists a local Hilbert space transformation in the lines exposed in \cite{Basa:2019ywr} for $T\bar{T}$-deformed CFTs in the perturbative regime captured by cMERA. 
 
We close this section with some comments on the $\mc{O}(\mu)$-correcting term for $S_A$ in \eqref{eq:ee_cmera_tt}. It is well-known that there is a relation  between the entanglement entropy and the conformal trace anomaly \cite{Holzhey:1994we}. Classically, the trace of the energy-momentum tensor vanishes as a consequence of conformal invariance. Nevertheless, the need for explicitly regulating the QFT brings a non-vanishing value for $\la T^{a}_a\ra \propto c\,R \neq0$, where $R$ is the curvature of the non-trivial background metric resulting from breaking scale invariance and $c$ is the central charge of the theory. This specifically accounts for the appearance of extra correlations in products of the energy-momentum tensor at short distances, and thus links the divergencies in the entanglement entropy with the short distance behavior of the QFT. A cMERA construction can be understood as an explicit regularization of a QFT through the cMERA cutoff. Thus, using our optimized cMERA solution for the $T\bar{T}$-deformed scalar theory and the $T\bar{T}$-trace relation \eqref{eq:tracerel}, at order $\mu$, we obtain: \beq\label{eq:tracerel_cmera}
\la T_a^a\ra_{\Lambda} \equiv 2\mu\, \la \Psi_{\Lambda}\, |\mc{O}^{(1)}_{T\bar{T}}|\, \Psi_{\Lambda}\ra = \frac{\widetilde{\alpha}}{12 \pi}\Lambda^{4}\mu\,
\eeq 
with $\widetilde{\alpha}=13-8 \sqrt{2}$. Recall that we use the magic cMERA realization to achieve a non-trivial state that is aimed to approximate the ground state of the $T\bar{T}$-deformed CFT. This corresponds to (\ref{eq:hamilt}), and thus has no UV-regulator in it. We move away from that theory by including a UV-regulator (modifying the UV-region) to obtain the desired approximating cMERA state. However, once we have obtained such state, if we want to compute correlators among operator that are sensitive to the particular theory one considers, such as stress-energy tensor correlators, we need to re-consider the original $T\bar{T}$-deformed theory (\ref{eq:hamilt}) (even though we are not minimizing such Hamiltonian). Then the trace-relation holds and we can use it, for instance, to find the expectation value of the trace of the stress-energy tensor as we just did.

It is now convenient to define a \emph{fictious} scalar curvature $R(\mu)$ as \beq\label{eq:ricci_curv}
R(\mu) \equiv \frac{1}{\widetilde{\alpha}\Lambda^{2}}\la T_a^a\ra_{\Lambda} =\frac{1}{12 \pi}\Lambda^{2}\mu\,,
\eeq
which encodes the appearance of extra correlations induced by the $T\bar{T}$ deformation in products of the energy-momentum tensor at short distances. The curvature sign depends on the sign of the deformation parameter $\mu$. As a result, one may trace back the correcting term in $S_A$ to the $T\bar{T}$-trace relation \eqref{eq:tracerel} since \beq\label{eq:ee_cmera_tt_ricci}
 S_A = \frac{\pi}{24}\Lambda\left(1 + \frac{\alpha}{\pi}R(\mu)\right) + \widetilde{\kappa}_{\Lambda}\, .
\eeq
 
\subsubsection{Holography: cMERA and the sharp radial cutoff proposal}\label{subsec:holo}
Recently, McGough, Mezei and Verlinde \cite{McGough:2016lol}  proposed, in the context of the AdS/CFT, a correspondence between the emergent holographic radial direction and the $T\bar{T}$ flow parameter $\mu$ in the dual CFT. Concretely, it has been conjectured that a $T\bar{T}$-deformed CFT with $\mu >0$ (in our convention) is dual to AdS$_3$ gravity with a sharp radial cutoff located at $r_c \sim 1/\sqrt{\mu}$ in the background of a BTZ black hole\footnote{In coordinates such that the boundary of AdS lies at $r \to \infty$.}\,. The proposal suggests that the bulk degrees of freedom outside the surface $r = r_c$ must be removed\footnote{The relation between a radial cutoff in the bulk and a UV cutoff in the boundary field theory has never been made precise, but it has been largely discussed in the context of the holographic RG \cite{deBoer:1999tgo, Heemskerk:2010hk, Faulkner:2010jy}. What is clear is that integrating out degrees of freedom above a particular UV scale in the boundary corresponds to integrating out the fluctuations of the bulk fields outside a given radius.}\,. It has been pointed out that the reduction in the number of degrees of freedom implicit in the sharp cutoff proposal would disrupt the integrability and the UV completeness of the deformed CFT. Therefore, it seems natural that the sharp radial cutoff proposal will need to be reconsidered \cite{Guica:2019nzm}. In this regard, the careful analysis of Ref.~\cite{McGough:2016lol} shows that the AdS dual of a $T\bar{T}$-deformed CFT represents a mild modification of AdS: it just amounts to a linear mixing of bulk fields that effectively reduce to Dirichlet boundary conditions at $r_c$.

Here we analyze our results in light of the conjectured correspondence between cMERA and holography \cite{Swingle:2009bg,Swingle:2012wq,Nozaki:2012zj}. Therefore, the discussion about the sharp radial cutoff becomes a discussion about the cutoff in the cMERA flow of the $T\bar{T}$-deformed theory.

First, recall that the half-space EE for this theory does not violate the area law of entanglement. Following the results and discussions stated in \cite{Fernandez-Melgarejo:2021ymz}, we note that a putative dual spacetime derived from the EE in Eq.~\eqref{eq:ee_cmera_tt} should be kept entire. In other words, the cMERA result does not hint that a region outside the cMERA ``cut off'' surface must be removed. To clarify this, let us dig in a little deeper into the holographic ``cut off'' surface analogue in this cMERA setup. When focusing on the cMERA flow starting at the IR disentangled state $|\Omega\ra$ and running up to scales $k \ll \Lambda$, the cMERA Boguliubov angle is effectively reduced to \eqref{eq:dispersion_relation_cl}. This simple form of the cMERA variational parameter can be casted as \beq
f(k,s_{\text{IR}}) = \frac{1}{2}\log\frac{\epsilon_0(k)}{\Lambda_{\beta}}\, ,\hspace{0.75cm} \Lambda_{\beta}\equiv \frac{\Lambda}{\beta(\mu)} \, ,
\eeq
with $\Lambda_{\beta}<\Lambda$ for $\mu >0$. Accordingly, one may state that there is a fixed scale, $-\infty<s_c<s_{\text{UV}}$, at which the cMERA representation of the $T\bar{T}$-deformed ground state simply amounts to the cMERA ground state of the undeformed scalar but defined on a renormalized momentum cutoff, $\Lambda_{\beta}$, with the hard cutoff function given in \eqref{eq:hard_cutoff}. Namely, the half-space EE for this state is merely $S_A(0)$, which shows no hint of the effect of the $T\bar{T}$ deformation. The scale $s_c$ at which the cMERA flow attains this state, $|\Psi_{\Lambda_{\beta}}\ra$, can be derived straightforwardly by noting that in cMERA any scale defines an effective cutoff through $\Lambda_s = \Lambda\, e^{s}$. In our case $\Lambda_{\beta} = \Lambda\, e^{s_c}$ and therefore \beq
s_c = -\log\beta(\mu) \simeq -\frac{a}{3\pi}\Lambda^{2}\mu\, .
\eeq
This indicates that, when considering the full cMERA flow from $s_{\text{UV}}$ to $s_{\text{IR}}$, the original free scalar theory resurfaces at $s_c$. However, according to the EE result, it is not necessary to cut off the cMERA from $s_{\text{UV}}$ to $s_c$ in order to obtain a well-defined geometric dual of the tensor network. In other words, since the $\mc{O}(\mu)$ contribution to the EE, which is related to the flow between $s_{\text{UV}}$ and $s_c$, does not violate the area law of entanglement, then there is no contradiction with it being related with a piece of a Ryu-Takayanagi minimal surface.

In addition, the authors of Refs.~\cite{Kruthoff:2020hsi} and \cite{Caputa:2020fbc} provided arguments linking continuous tensor networks to the AdS$_3/$CFT$_2$ correspondence, where the field theory includes a $T\bar{T}$ deformation. In this framework, the ground states of an undeformed CFT and the deformed counterpart are related by a unitary operator representing the flow along $\mu$. It is expected that those ground states would be captured equivalently in terms of a continuos tensor network mapping. While finding this tensor network mapping for a finite (non-perturbative) evolution in $\mu$ is a challenging problem \cite{Kruthoff:2020hsi}, our setting allows us to directly find a perturbative solution. To see this, consider a unitary operator that maps the two ground states of the theories under consideration, i.e.~the ground state of the free scalar theory $|\Psi_{\Lambda}(0)\ra$ and the ground state of the deformed theory $|\Psi_\Lambda(\mu)\ra$,\beq\label{eq:TT_unitary_gen}
|\Psi_\Lambda(\mu)\ra = \mathbb{U}_{T\bar{T}}\, |\Psi_{\Lambda}(0)\ra \, .
\eeq
In this equation both states are Gaussian states defined through their respective variational Bogoliubov angles. The unitary operator connecting both states is given by \cite{Cotler:2018ehb,Cotler:2018ufx}
\beq\begin{split}
\mathbb{U}_{T\bar{T}} &= \exp\left(-\frac{i}{4} \int \dd k\, \log\left(\frac{\epsilon_0(k)}{\epsilon(k)}\right)\chi_{T\bar{T}}(k)\right) \\
&= \exp\left(\frac{i}{4} \int \dd k\, \log \beta(k;\mu)\chi_{T\bar{T}}(k)\right),
\end{split}\eeq
where $\chi_{T\bar{T}}(k)$ is the quadratic operator $\chi_{T\bar{T}}(k)= \phi(k) \pi(-k) + \pi(k)\phi(-k)$. Assuming that $\Lambda^{2}\mu$ is small enough, \beq
\log \beta(k;\mu) \simeq \frac{a}{3 \pi}\left(\frac{bk^2 +\Lambda^2}{k^2+\Lambda^2}\right)\Lambda^2\mu\, ,
\eeq
and it is then possible to define the infinitesimal $T\bar{T}$-flow addressed in this paper in terms of a Gaussian cMERA-like unitary operator. This is given by \beq\begin{split}\label{eq:cmera-like-lambda-ev}
\mathbb{U}_{T\bar{T}} &= e^{-i\delta w\, \mc{K}_{T\bar{T}}} \\
&= \exp\left(-\frac{i}{2}\delta w \int \dd k\, g_{T\bar{T}}(k)\chi_{T\bar{T}}(k)\right),
\end{split}\eeq
where\beq
\mc{K}_{T\bar{T}} = \frac{1}{2}\int \dd k\, g_{T\bar{T}}(k)\chi_{T\bar{T}}(k)\, ,
\eeq 
and\beq
\delta w \equiv s_c= -\frac{a}{3 \pi}\Lambda^{2}\mu\, , \hs{0.75}  g_{T\bar{T}}(k)=\frac{1}{2}\frac{bk^{2} + \Lambda^{2}}{k^{2} + \Lambda^{2}}\,.
\eeq
Hence, one may write the $\mu$-flow in terms of a Gaussian cMERA evolution in a new scale $w\propto \mu$, at least infinitesimally. A remarkable property of \eqref{eq:cmera-like-lambda-ev} is that, within this approximation, the deformation merely acts by implementing a canonical/Bogoliubov transformation along the flow.  This implies that, on the line, the $\mu$-deformed state obtained through the cMERA-like evolution in $w$ retains its symmetries (including conformal symmetry), as the undeformed state \cite{Hu:2017rsp}. Notice that the $\mu$-evolution cMERA-like operator $\mathbb{U}_{T\bar{T}}$ represents a transformation of the undeformed state $|\Psi_{\Lambda}(0)\ra$ that is well defined at all length scales.

That said, it is expected that at finite $\mu$ one could presumably write the evolution as \beq
\mathbb{U}_{T\bar{T}} = \exp\bigg(-\frac{i}{2}\int_0^{w(\mu)}\dd w\int \dd k\, g_{T\bar{T}}(k;\mu)\chi_{T\bar{T}}(k;\mu)\bigg),
\eeq
where the entangling operator $\chi_{T\bar{T}}(k;\mu)$ must non-perturbatively include non-quadratic higher-order operators. While challenging, the results in \cite{Kruthoff:2020hsi} suggest a way for building up this setup. This is crucial in order to elucidate if it is possible to adscribe a sensible geometric dual to the finite evolution in $w$ in the lines exposed in \cite{Caputa:2020fbc}.

\section{Conclusions and Outlook}\label{sec:concl}
In this work we obtain and analyze the RG flow of a $T\bar{T}$-deformed scalar CFT on the line and to first order in the deformation parameter through a (Gaussian) cMERA continuous tensor network. Our approach provides a concrete setting to study non-local effects in quantum field theory from a tensor network perspective. 

Interestingly, we found that cMERA, when treating an irrelevant deformation such as the $T\bar{T}$, shows a non-trivial sensitivity to concrete realizations of its variational parameter. The construction gives, via the minimization of the energy functional on the variational wave-function, a density of disentanglers from which one can read the deformed dispersion relation. This result allows us to find the correction to the correlators of scaling operators of the theory. Then we compute the entanglement entropy in order to analyze the nature of the non-localities induced by the $T\bar{T}$ deformation at short length scales. Within the scope of our approximation, we find that the kind of non-locality generated by the $T\bar{T}$ deformation can be regarded as a mild-one, in the sense that it does not violate the area law of entanglement, which holds for the ground states of local Hamiltonians. In the context of the conjectured connection between cMERA and holography, we find that when the deformation parameter is positive, a finite bulk radius in the putative geometric description of the optimized cMERA solution can be defined. In our setting, this landmarks the scale at which the deformed state becomes the original one. Nevertheless, the entropy analysis contradicts the proposal that no geometry can be ascribed to the region outside this radial cutoff.

It is worth to notice that most remarkable properties of the $T\bar{T}$ deformation  such as the modification of the energy levels, manifest when considering theories on the cylinder, that is, when the spatial coordinate is compactified on a circle \cite{Zamolodchikov:2004ce, Smirnov:2016lqw, Cavaglia:2016oda}. One may ask whether and how our analysis on the line, can be extended to this situation. It turns out that the MERA formalism in the circle has been recently proposed \cite{Hung:2021tsu}. This generalization consists in wrapping the action of the optimized entangler in the line around the circle. The wrapping does not affect the operator content of the entangler but only the structure of the variational parameter $g(k;s)$. Specifically, in the case of a Gaussian cMERA, the wrapping of the entangler around the circle may be implemented through the method of images yielding an important and practical result: if a cMERA state on the line represents a good approximation to a ground state of a local Hamiltonian, then the resulting cMERA on a circle is also a good approximation to the ground state of the same local Hamiltonian, but now defined on the circle. We leave for future investigations to analyze if our solution for the $T\bar{T}$-deformed free scalar in the circle would provide additional insights on the structure of entanglement and the correlation functions of a $T\bar{T}$-deformed CFT, as well as in their holographic interpretations.

\section*{Acknowledgments}
We thank J.J.~Fern\'andez Melgarejo for very fruitful discussions and a careful reading that helped to improve the manuscript. B.C.~is financially supported by Fundaci\'on S\'eneca de la Regi\'on de Murcia under the grant 35642/POST/19. B.C.~is also supported by a Margarita Salas Postdoctoral Research Fellowship funded by Next Generation EU. J.M.V.~is funded by Ministerio de Ciencia, Innovaci\'on y Universidades PGC2018-097328-B-100 and Programa de Excelencia de la Fundaci\'on S\'eneca de la Regi\'on de Murcia 19882/GERM/15. 

\begin{appendix}
\section{Energy functional on cMERA}\label{app:1}
In this appendix section we compute the energy expectation value on the family class of gaussian cMERA states. The latter is given by (\ref{eq:cMERAstate}), and the Hamiltonian we work with is (\ref{eq:h0hTT}). Generically, the expectation value of a local operator $\mc{O}(x)$ on cMERA is given by:\footnote{For simplicity, in this section we will obviate the label cMERA to point out the state in which the expectation value is computed. It is assumed that expectation values are evaluated on such state.} \beq\begin{split}
 \la\mc{O}\ra_{\text{cMERA}} &= \la\Omega|U^\dag(s_{\text{IR}},s)\,\mc{O}\,U(s_{\text{IR}},s)|\Omega\ra \\
 &= \la\Omega|\mc{O}(s_{\text{IR}})|\Omega\ra\,,
\end{split}\eeq
where \beq
\mc{O}(s_{\text{IR}}) = \int\dd x\, \mc{O}(x;s_{\text{IR}})\,.
\eeq

To proceed it is convenient to work in momentum space, where the canonical variables decompose as \beq
\phi(x) = \int\frac{\dd k}{2\pi}\phi(k)e^{ikx}\,, \hs{0.75} \pi(x) = \int\frac{\dd k}{2\pi}\pi(k)e^{ikx}\,.
\eeq
Then the Hamiltonian (\ref{eq:h0hTT}) to first order in $\mu$ reads: \beq\begin{split}
H_\Lambda &= \frac{1}{2}\int\frac{\dd k}{2\pi}\Big(\pi(k)\pi(-k) + k^2\phi(k)\phi(-k) + \frac{k^2}{\Lambda^2}\pi(k)\pi(-k)\Big) \\
&\hs{0.45}+\frac{\mu}{4}\int\frac{\dd k_1\dd k_2\dd k_3}{(2\pi)^3} \Big(\pi(k_1)\pi(k_2)\pi(k_3)\pi(k_4) +k_1k_2k_3k_4\phi(k_1)\phi(k_2)\phi(k_3)\phi(k_4) \\
&\hs{3.75}+2k_3k_4\pi(k_1)\pi(k_2)\phi(k_3)\phi(k_4)\Big), \\ 
\end{split}\eeq
where in the last two lines $k_4 = -k_1-k_2-k_3$. From this expression one can easily identify the free Hamiltonian with the UV-regulator term (\ref{eq:UVreg}) and the leading-order contribution (\ref{eq:OTTmu}), which in terms of canonical fields reads: \beq
\mc{O}^{(1)}_{T\bar{T}}(x) = \frac{1}{4}\Big(\pi^4 + (\phi')^4 - 2\pi^2(\phi')^2\Big).
\eeq
At this point, evaluate on the cMERA state is equivalent to substitute $\phi(k)$ and $\pi(k)$ by their scale-dependent counterparts given in (\ref{eq:canotransf}). Then we get the expectation value just by projecting-out on the (reference) vacuum state $|\Omega\ra$.

For instance, for the free Hamiltonian with the regulator term we have:
\beq
\la\bar{\mc{E}}_{0}\ra[f] = \frac{\pi\Lambda}{2}\int\frac{\dd k}{2\pi}\Big(e^{2f(k,s_{\text{IR}})}\Big(1+\frac{k^2}{\Lambda^2}\Big) + \frac{k^2}{\Lambda^2}e^{-2f(k,s_{\text{IR}})}\Big)\delta(0)\,.
\eeq
where we have used the 2-point functions: \beq\begin{split}
\la\Omega|\phi(p)\phi(q)|\Omega\ra &= \frac{\pi}{\Lambda}\delta(p+q)\,, \\
\la\Omega|\pi(p)\pi(q)|\Omega\ra &= \pi\Lambda\delta(p+q)\,, \\
\la\Omega|\pi(p)\phi(q)|\Omega\ra &= -i\pi\delta(p+q)\,.
\end{split}\eeq
For the leading order term the procedure is the same but it depends on 4-point instead of 2-point functions. Wick's theorem allows us to decompose them in terms of 2-point functions, obtaining combinations of products of two Dirac's delta. One can use one of them to perform an integral over momentum. The final result for this piece is: \beq\begin{split}\label{Ott1}
\la\mc{O}^{(1)}_{T\bar{T}}\ra[f] &= \frac{\pi\Lambda^2}{8}\int\frac{\dd k}{2\pi}\frac{\dd k_1}{2\pi}\Big(3e^{2(f(k,s_{\text{IR}})+f(k_1,s_{\text{IR}}))}+3\frac{k^2k_1^2}{\Lambda^4}e^{-2(f(k,s_{\text{IR}})+f(k_1,s_{\text{IR}}))} \\
&\hs{3.2}-2\frac{k_1^2}{\Lambda^2}e^{2(f(k,s_{\text{IR}})-f(k_1,s_{\text{IR}}))} + 4\frac{kk_1}{\Lambda^2}\Big)\delta(0)\,,
\end{split}\eeq
The desired functional is then given by \beq
\la\mc{E}\ra[f] = \la\bar{\mc{E}}_{0}\ra + \mu\la\mc{O}^{(1)}_{T\bar{T}}\ra + \mc{O}(\mu^2)\,.
\eeq
Upon minimization $\delta\la\mc{E}\ra[f]/\delta f(k,s_{\text{IR}}) = 0$, one obtains: \beq
\label{eq:gap_equation}
e^{2f(k,s_{\text{IR}})} = \frac{|k|}{\Lambda\sqrt{1+\left(k/\Lambda\right)^{2}}}+\mu\frac{4 \Lambda ^2 k(I_{(2)}^--\Lambda ^2I_{(0)}^+)+k^3(3I_{(2)}^--\Lambda ^2I_{(0)}^+)}{4 \Lambda^4\left(1+\left(k/\Lambda\right)^{2}\right)^{3/2}}\,,
\eeq
where $I_{(0)}^{+}$ and $I_{(2)}^{-}$ are defined as: \beq
I_{(n)}^\pm = \int \frac{\dd p}{2\pi}\,p^n e^{\pm2f(p,s_{IR})}\,.
\eeq

Notice that Eq.~\eqref{eq:gap_equation} defines an integral equation for $f(k,s_{\text{IR}})$ and must be self-consistently solved. The integrals $I_{(0)}^{+}$ and $I_{(2)}^{-}$ can be evaluated iteratively by taking as starting point $f_0(k,s_{\text{IR}}) = f(k,s_{\text{IR}})\big|_{\mu = 0}$, i.e.~the Gaussian state corresponding to the undeformed theory (now including the UV-regulator $\Gamma_{\text{UV}}$). At the first iteration, the integrals $I_{(0)}^{+}$ and $I_{(2)}^{-}$ turn out to be: \beq
I_{(2)}^- = \frac{2\sqrt{2}-1}{3\pi}\Lambda^3\,, \hs{0.75} I_{(0)}^+ = \frac{\sqrt{2}-1}{\pi}\Lambda\,.
\eeq
Plugging these expressions back on to (\ref{eq:gap_equation}), one obtains (\ref{eq:f_bogoliubov}).

\subsection{2nd order w/sharp cutoff}\label{subsec:appA1}
In section \ref{subsec:opt} we claim that the optimization of the expectation value of the Hamiltonian using the sharp cutoff cMERA scheme, which amounts to consider as 0th order Hamiltonian $H_0$ (with no UV-regulator), also vanishes at 2nd order in perturbation theory. In this section we proof that claim.

The 2nd order $T\bar{T}$ operator is given by:
\beq
\mc{O}_{T\bar{T}}^{(2)}(x) = 2\mc{H}_0\mc{O}_{T\bar{T}}^{(1)}\,.
\eeq
In terms of canonical fields, this corresponds to the operator: \beq
\mc{O}_{T\bar{T}}^{(2)}(x) = \frac{1}{4}\Big(\pi^6 + (\phi')^6 - \pi^4(\phi')^2 - \pi^2(\phi')^4\Big).
\eeq
Its expectation value on cMERA is: \beq\begin{split}\label{Ott2}
\la\mc{O}^{(2)}_{T\bar{T}}\ra[f] &= \frac{\pi}{16}\int\frac{\dd k}{2\pi}\frac{\dd k_2}{2\pi}\frac{\dd k_3}{2\pi}\Big(15\Lambda^3e^{2\left(f(k,s_{\text{IR}})+f(k_2,s_{\text{IR}})+f(k_3,s_{\text{IR}})\right)} \\
&\hs{3.65}+\frac{15}{\Lambda^3}k^2k_2^2k_3^2e^{-2\left(f(k,s_{\text{IR}})+f(k_2,s_{\text{IR}})+f(k_3,s_{\text{IR}})\right)} \\
&\hs{3.65}-3\Lambda e^{2f(k,s_{\text{IR}})}\left(k_3^2e^{2(f(k_2,s_{\text{IR}})-f(k_3,s_{\text{IR}}))} - 4 k_2k_3\right) \\
&\hs{3.65}-\frac{3}{\Lambda}k_2^2k_3^2e^{2(f(k,s_{\text{IR}})-f(k_2,s_{\text{IR}})-f(k_3,s_{\text{IR}}))}\Big)\delta(0)\,.
\end{split}\eeq
The desired functional is now given by \beq
\la\mc{E}\ra[f] = \la\mc{E}_{0}\ra + \mu\la\mc{O}^{(1)}_{T\bar{T}}\ra + \mu^2\la\mc{O}^{(2)}_{T\bar{T}}\ra + \mc{O}(\mu^3)\,,
\eeq
where $\la\mc{E}_{0}\ra = \la\bar{\mc{E}}_{0} - \Gamma_{\text{UV}}^\Lambda\ra$, and $\la\mc{O}^{(1)}_{T\bar{T}}\ra$ and $\la\mc{O}^{(2)}_{T\bar{T}}\ra$ are given by (\ref{Ott1}) and (\ref{Ott2}) respectively. Repeating the optimization procedure as before, one obtains: \beq
f(k,s_{\text{IR}}) = \frac{1}{2}\log\frac{|k|}{\Lambda}\Bigg(1+\left(\frac{1}{\Lambda}I_{(2)}^- - \Lambda I_{(0)}^+\right)\left(\mu + \mu^2\Big(\frac{3}{\Lambda}I^-_{(2)}+2\Lambda I_{(0)}^+\Big)\right)\Bigg),
\eeq
which completes Eq.~(\ref{eq:eqEeq1}). This expression becomes independent of $\mu$ if we evaluate the integrals for the seed theory (i.e.~at the first iteration, where $I_{(2)}^- = \Lambda^2 I_{(0)}^+$).

\end{appendix}

\small

\bibliography{references}
\bibliographystyle{utphys}

\end{document}